\documentclass[3p,times,procedia]{elsarticle}
\usepackage{nupha_ecrc}
\usepackage{wrapfig}

\volume{00}

\firstpage{1}

\journalname{Nuclear Physics A}

\runauth{Roy A. Lacey for the STAR Collaboration}

\jid{nupha}

\jnltitlelogo{Nuclear Physics A}

\usepackage{amssymb}
\usepackage[figuresright]{rotating}
\newcommand{\dphi}{\Delta\phi}
\newcommand{\Yphi}{$Y(\Delta\phi)$}
\newcommand{\YphiTempl}{Y(\Delta\phi)^{templ}}
\newcommand{\YphiRidge}{Y(\Delta\phi)^{ridge}}

\newcommand{\Yphipp}{Y(\Delta\phi)^{pp}}

\begin{document}

\begin{frontmatter}



%% Instructions from Editor: Please use the following \dochead only in the preprint version (e-print arXiv etc.); 
%% use empty \dochead{} when submitting to Nuclear Physics A!
\dochead{XXVIIIth International Conference on Ultrarelativistic Nucleus-Nucleus Collisions\\ (Quark Matter 2019)}
%% e.g. \dochead{17th International Conference on Dynamical Processes in Excited States of Solids}

\title{Long-range collectivity in small collision-systems with \\ two- and four-particle correlations @ STAR}


\author{Roy A. Lacey for the STAR Collaboration}

\address{Depts. of Chemistry \& Physics, Stony Brook University, Stony Brook, NY, USA}

\begin{abstract}
New STAR  differential  and integral $v_{2,3}$ measurements that explicitly account for non-flow contributions, 
are reported  for $p$/$d$/$^3$He+Au, collisions at $\sqrt{s_{_{NN}}}=200$~GeV.
The measurements, which leverage the two-particle correlators for $p$/$d$/$^3$He+Au  and 
minimum-bias $p+p$ collisions in tandem with three well established methods of non-flow subtraction, are 
observed to be method-independent. For comparable multiplicities, they further indicate  
system-independent $v_2\{2\}$ and $v_3\{2\}$ values that are consistent with the critical role of 
both “size” ($\rm{N_{ch}}$) and the subnucleonic-fluctuations-driven eccentricities $\varepsilon_{2,3}$, but 
are inconsistent with the notion of shape engineering in $p$/$d$/$^3$He+Au collisions.
The scaling function derived from the measurements, confirm the important role of final-state effects 
across a broad spectrum of collision-system sizes and energies, and suggests 
 an increase in $\eta/s(T,\mu_B)$ for small collision-systems.

\end{abstract}

\end{frontmatter}

% \linenumbers

\section{Introduction}
\label{intro}
Relativistic heavy-ion collisions can lead to  
high energy density strongly interacting matter with an anisotropic 
transverse energy density profile. This matter not only quenches jets, but  also 
expands and hadronize to produce particles with an azimuthal anisotropy that reflects 
the viscous hydrodynamic response to the initial energy density profile 
\cite{Song:2010mg}.
%Teaney:2003kp,
%Romatschke:2007mq,Song:2010mg,Staig:2011wj,C,
%Bozek:2011if,Gardim:2012yp,Shuryak:2013ke,Qian:2016fpi,McDonald:2016vlt,Bernhard:2016tnd}. 
%
The shape of this profile, $\rho_e(r,\varphi)$, can be characterized 
by the complex eccentricity vectors \cite{Teaney:2010vd,Qiu:2011iv}:
%Alver:2010dn,Petersen:2010cw,Lacey:2010hw,
%
%\begin{eqnarray}
$
{\mathcal{E}_n  \equiv \varepsilon_n e^{in\Phi_n} \equiv 
  - {\int d^2r_\perp\, r^m\,e^{in\varphi}\, \rho_e(r,\varphi)}/
           {\int d^2r_\perp\, r^m\,\rho_e(r,\varphi)}},                                                       
%\label{epsdef1}
%\end{eqnarray}
$
where ${\varepsilon_n = {\left< \left| \mathcal{E}_n \right|^2 \right>}^{1/2}}$ and ${\Phi_n}$ denote 
the magnitude and azimuthal direction of the $\mathrm{n^{th}}$-order eccentricity vector 
which fluctuates from event to event \cite{Teaney:2010vd}. 
The eccentricity fluctuations are driven by both nucleonic and subnucleonic fluctuations and can be 
estimated via a quark Glauber model.
%,Bhalerao:2014xra,Yan:2015jma
%
The quenched jets and the anisotropic flow which derives from the pressure gradients induced by $\mathrm{\varepsilon_n}$, 
result in an azimuthal anisotropy of the measured single-particle distribution, quantified 
by the complex anisotropy vectors \cite{Qiu:2011iv}:
%,Luzum:2011mm,Teaney:2012ke,Ollitrault:1992bk,
%
$V_n  \equiv v_ne^{in\Psi_n}$,  
$\equiv \{e^{in\phi}\}$, ${v_n = {\left< \left| V_n \right|^2 \right>}^{1/2}}$,
%\label{Vndef}
%\end{equation}
%
where $\phi$ denotes the azimuthal angle around the beam direction, of a particle emitted 
in the collision, $\{\dots\}$ denotes the average over all particles emitted in the event, 
and $\mathrm{v_n}$ and $\mathrm{\Psi_n}$ denote 
the magnitude and azimuthal direction of the $\mathrm{n^{th}}$-order anisotropy vector 
which also fluctuates from event to event. 
%The coefficients $\mathrm{v_2}$ and $\mathrm{v_3}$
%are termed elliptic- and triangular flow respectively.
Model comparisons to $\mathrm{v_n}$ measurements continue to be an important avenue to estimate  
the transport coefficients for the partonic matter produced in large to moderate-sized 
collision systems \cite{Song:2010mg,Qiu:2011iv,Schenke:2011tv}.  
%\cite{Hirano:2005xf,Romatschke:2007mq,Song:2010mg,Schenke:2010rr,Bozek:2010wt,
%Qiu:2011iv,Schenke:2011tv,Niemi:2012ry,Gardim:2012yp,McDonald:2016vlt,Bernhard:2016tnd} for the 

For the small collision-systems produced in $p$/$d$/$^3$He+Au and $p$+Pb collisions, collective flow might 
not develop due to the presence of large gradients that could 
excite non-hydrodynamic modes or render invalid, the hydrodynamic 
gradient expansion  \cite{Denicol:2012cn,Florkowski:2016zsi} required 
%,
to accurately characterize the viscous hydrodynamic response. Indeed, a most vexing question that
pervades our field is whether an alternative initial-state-driven mechanism \cite{Dusling:2012iga} 
prevails over hydrodynamic expansion for these collision-systems.
However,  numerical simulations in strongly interacting theories suggest that hydrodynamics remains 
applicable even when the system size ($\mathrm{R}$) is of ${\cal O}(1/\mathrm{T})$ -- the inverse 
temperature  \cite{Chesler:2016ceu}. Here, subnucleonic fluctuations become crucial.
%That is, when the dimensionless size $\mathrm{RT} \sim 1$ \cite{Chesler:2016ceu}.

The current measurements for $p$/$d$/$^3$He+Au collisions, which supplement earlier 
measurements at both RHIC \cite{PHENIX:2018lia} and 
%Adare:2015ctn
the LHC \cite{Chatrchyan:2013nka}
%,Abelev:2012olaAad:2012gla,Aaboud:2017acw
aim to address the respective influence of collision-system size, $\varepsilon_n$ and its attendant 
subnucleonic fluctuations and viscous attenuation on the measured non-flow-mitigated $v_n$.

\section{ Two particle correlators and $v_n$ extraction}
\begin{figure}[htb]
%\begin{minipage}{0.97\linewidth}
%\begin{figure}[h]
\includegraphics[width=1\linewidth]{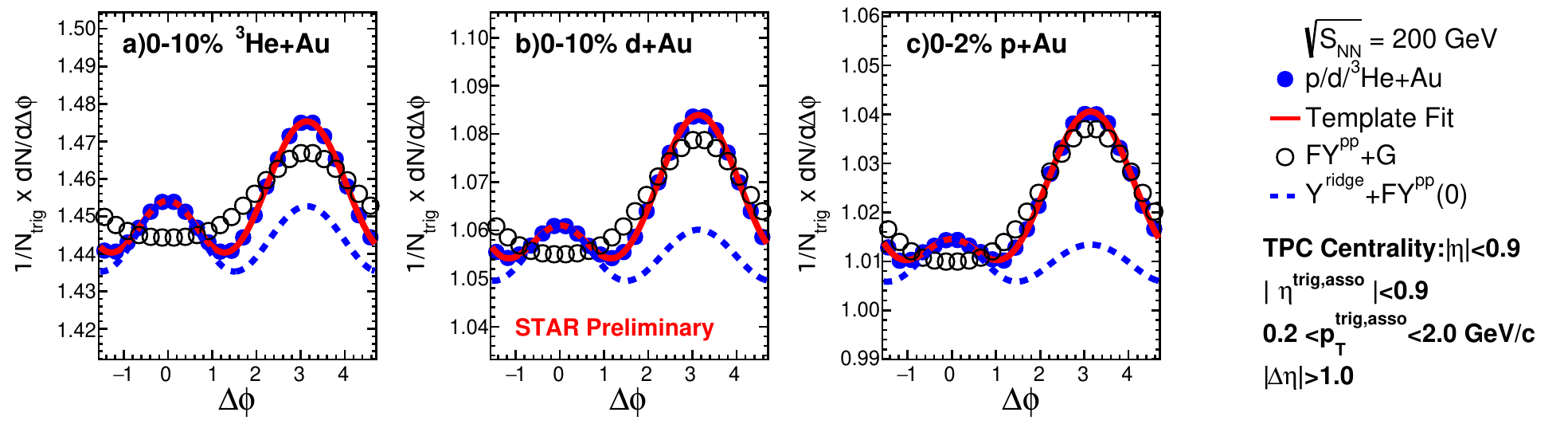}
\vskip -0.35cm
\caption{ Illustration of the template fitting procedure which employs the MB $p+p$ correlator to estimate 
the non-flow contributions and to extract the $v_2$ and $v_3$ Fourier coefficients. The $p_T$ and $\Delta\eta$ selections for the 
correlation functions are as indicated.} 
\label{qm:tempall}
%\end{minipage}
\end{figure}

%
%%
%\begin{figure}[bh]
%%\begin{minipage}{0.97\linewidth}
%%\centering
%\includegraphics[width=1.0\linewidth]{QM_Four_All.pdf}
%\vskip -0.35cm
%\caption{Two-particle per-trigger yield distributions for for $^3$He/d/p+Au and \pp\ collisions at \TopE\ as 
%indicated; the trigger and associated particles are both selected from the range $0.2 <p_{T}< 2.0$~GeV. 
%}
%\label{qm:fourall}
%\end{figure}
%\end{minipage}
%
%\end{figure}
%
The measurements were obtained with the STAR detector, via the two-particle correlation method.
The per-trigger yields  \Yphi\ = $1/N_{\rm Trig}*dN/d\Delta\phi$  for 
$0-2\%$ $p$+Au, $0-10\%$ $d$+Au and $0-10\%$ $^{3}$He+Au collisions are shown as a function 
of $\dphi$ in Figs.~\ref{qm:tempall} (a)-(c); for these correlators, the trigger ($\rm Trig-$) and the 
associated ($\rm Assoc-$) particles are measured in the 
ranges $0.2 < p_{T} < 2.0$~GeV/c and $|\eta|< 0.9$. The requirement $|\Delta\eta|>1.0$ 
between the $\rm Trig$- and the $\rm Assoc-$particles was also imposed  to suppress possible non-flow 
contributions from the near-side jet. 
Figures~\ref{qm:tempall} (a)-(c) indicate a near-side ``ridge" suggestive of an influence from flow-like
 contributions to the measured correlators for $p$/$d$/$^{3}$He+Au collisions. The absence of this ridge
for minimum-bias (MB) $p+p$ collisions (c.f Fig.~\ref{qm:tempall}), further suggests that the $p+p$ 
correlator can be leveraged to obtain quantitative estimates of the non-flow contributions to 
the $p$/$d$/$^3$He+Au correlators.

Three separate methods were utilized to estimate and subtract the non-flow contributions to the measured differential 
correlation functions $Y(\Delta\phi, p_T, \rm{cent})$ used to extract $v_{2,3}(p_T)$ and $v_{2,3}(\rm{N_{chg}})$.
One is based on the template-fit method \cite{Aad:2016}.
The other two are based on Fourier expansion fits to the measured correlators \cite{adams:2004ja}. 
% Adams2005:aj,PhysRevC.98.014912
The template fitting procedure~\cite{Aad:2016} is illustrated in Fig.~\ref{qm:tempall}.
It assumes that the central $p$/$d$/$^{3}$He+Au \Yphi\ distributions are a superposition of 
a scaled MB \Yphi\ distribution for  $p+p$  collisions and a constant 
modulated by the ridge $\sum_{n=2}^{4}\ c_{n}^{sub}\cos(n\dphi)$ as: 
%shown in Eq.~\ref{eq:template}:
%
$
%\begin{equation}
\YphiTempl = F\Yphipp +  \YphiRidge\, ,
\label{eq:template}
%\end{equation}
$
where 
$
%\begin{equation}
\YphiRidge = G \left(1 + 2\sum_{n=2}^{4}\ c_{n}^{sub,sys} \cos{(n\Delta \phi)}\right)\, ,
\label{eq:template_ridge}
%\end{equation}
$
with free parameters $F$ and $c_{n}^{sub}$. The coefficient $G$, which
represents the magnitude of the combinatoric component of $\YphiRidge$, is fixed by requiring that
$\int_0^{\pi}{\mathrm{d}\dphi}\; Y^{\mathrm{templ}} = \int_0^{\pi}{\mathrm{d}\dphi} \;
Y^{\mathrm{HM}}$.  Fig. \ref{qm:tempall} shows that the template fit accounts for the data quite well.

%
%\begin{wrapfigure}{R}{0.75\linewidth}
\begin{figure}
  %\begin{center}
	\centering
\includegraphics[width=0.62\linewidth]{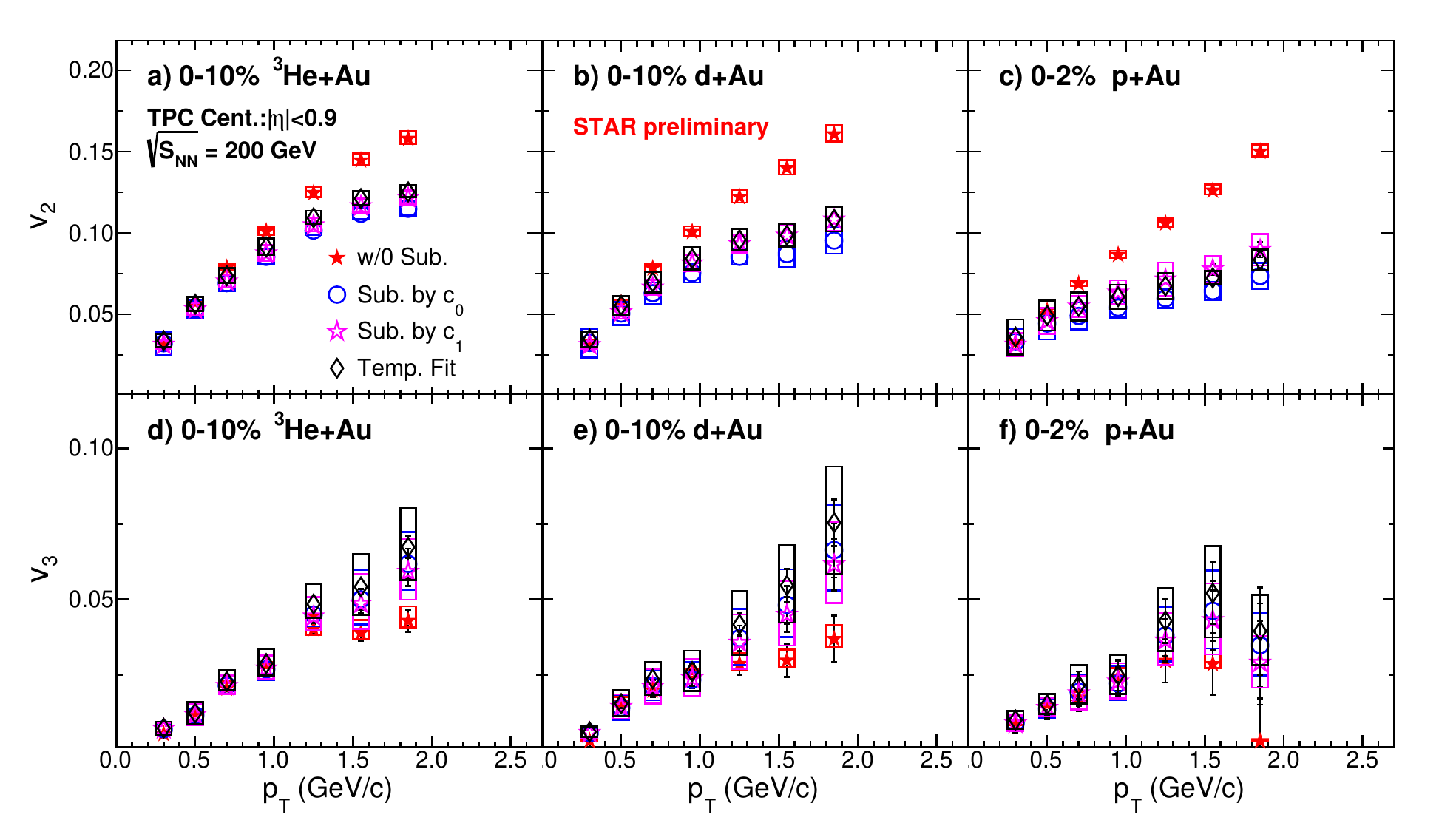}
\includegraphics[width=0.37\linewidth]{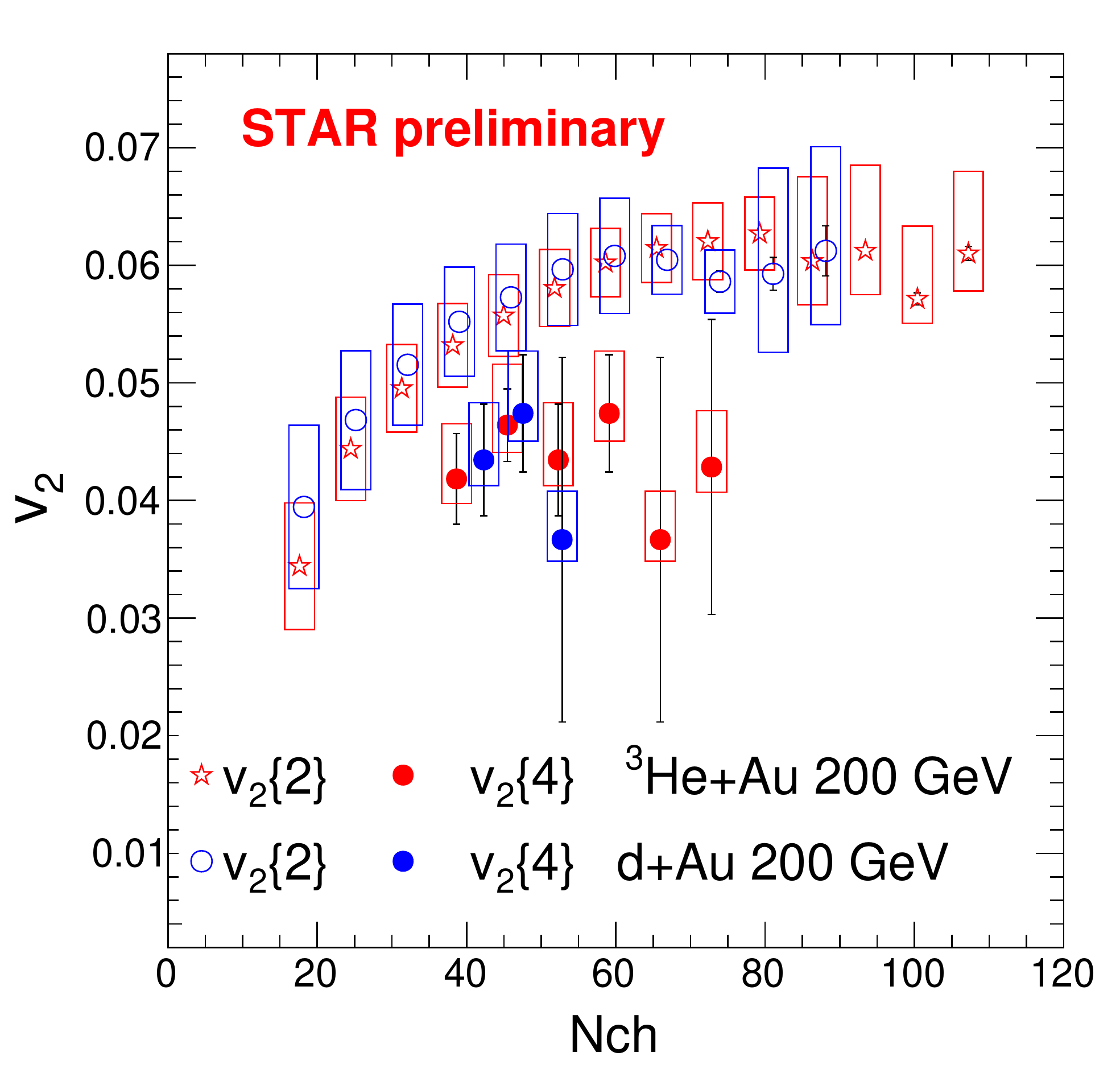}
	\vskip -0.35cm
  %\end{center}
 \caption{
Comparison of the $v_{2,3}(p_T)$  values for $p$/$d$/$^3$He+Au collisions,  
before and after non-flow subtraction, for all three methods (left panel). The right panel compares representative
$p_T$-integrated results for  $v_2\{2\}$ and $v_2\{4\}$ for 0-10\% $d$/$^3$He+Au collisions.
}
  \label{fig:method}
%\end{wrapfigure}
\end{figure}
Methods two and three rely on Fourier fits to the measured two particle correlators;
$
 %\begin{eqnarray}
 Y(\Delta\phi) = c_{0}(1 + \sum_{n=1}^{4} \, 2c_{n}\, \cos ( n \,\Delta\phi )). 
 \label{eq:fourier}
%\end{eqnarray}
$
Method two assumes that the non-flow contributions to $p$/$d$/$^3$He+Au is a superposition
of several proton-proton collisions. This leads to non-flow contributions that are equal 
to $c_2^{pp}$, but diluted by the pair-yield coefficient ($c_{0}$) difference between $p+p$ and $p$/$d$/$^3$He+Au.
%this method is similar to that for the ``scalar product method'' outlined in Refs.~\cite{adams:2004ja} and \cite{Adams2005:aj}). 
The subtracted coefficients $c_{n}^{sub}$ for $p$/$d$/$^3$He+Au are then obtained as:
$
%\begin{eqnarray}
c_{n}^{sub}  =  c_{n} - c_{n}^{pp}\times \frac{c_{0}^{pp}}{c_{0}},
\label{eq:c2_collective_noncollective}
%\end{eqnarray}
$
where $c_{n}=v_{n}^{\rm Trig} \times v_{n}^{\rm Assoc}$ -- the product of the flow coefficients $v_n$ 
for trigger- and associated-particles. Then $v_{n}^{\rm Trig}=c_{n}/v_{n}^{\rm Assoc}$ and 
$v_{n}^{\rm sub, Trig}=c_{n}^{sub}/v_{n}^{\rm sub, Assoc}$. 
Method three assumes that $c_{1}$ is dominated by the away-side jet. This leads to the estimate 
that the ratio of the non-flow between $p+p$ and $p$/$d$/$^3$He+Au is proportional to the ratio of 
the $c_{1}$ values for $p+p$ and $p$/$d$/$^3$He+Au respectively.  Thus, $c_{n}^{sub}$ can be obtained as:
$
%\begin{eqnarray}
c_{n}^{sub}  =  c_{n} - c_{n}^{pp}\times \frac{c_{1}}{c_{1}^{pp}}, 
%\label{eq:c2_collective_noncollective_2}
%\end{eqnarray}
$
and used to extract $v_n^{\rm sub}$ as described for method two.
It is noteworthy that closure tests were performed with simulated events from the AMPT model to 
aid validation of the efficacy of the respective methods for non-flow mitigation.

\section{Results}
%\subsection{$v_{n}$ in central p+Au, d+Au and $^{3}$He+Au at 200 GeV}
The $v_2(p_T)$ and $v_3(p_T)$ values for $p$/$d$/$^3$He+Au before and after non-flow subtraction, 
are compared for all three methods in the left panel of Fig.~\ref{fig:method}. They indicate 
non-flow contributions that are system-dependent, but the non-flow mitigated  $v_2(p_T)$ (top panels) and 
$v_3(p_T)$ (bottom panels) are method-independent within the indicated uncertainties. 
Here, it is noteworthy that the un-subtracted $v_3(p_T)$ is a lower limit since non-flow subtraction leads to higher 
$v_3(p_T)$ values. The uncertainties for $v_2(p_T)$ and $v_3(p_T)$ reflect statistical, as well as systematic 
uncertainties linked to  (i) track related backgrounds, (ii) pileup effects and (iii) the methods of non-flow subtraction. 
The right panel of Fig.~\ref{fig:method} indicates magnitudes and trends 
for the $p_T$-integrated $v_2\{2\}$ and $v_2\{4\}$ for $d$+Au and $^{3}$He+Au, that are consistent with an 
important influence from both subnucleonic eccentricity fluctuations and size-driven ($\rm N_{ch}$) viscous attenuation.
Note that the statistics available for the $p$+Au data precluded a statistically significant measurement of $c_2\{4\}$
and hence, $v_2\{4\}$.

%\subsection{Comparison of  STAR and PHENIX data}
%
%
\begin{figure}
%\begin{wrapfigure}{R}{0.80\linewidth}
	\centering
\includegraphics[width=0.49\linewidth]{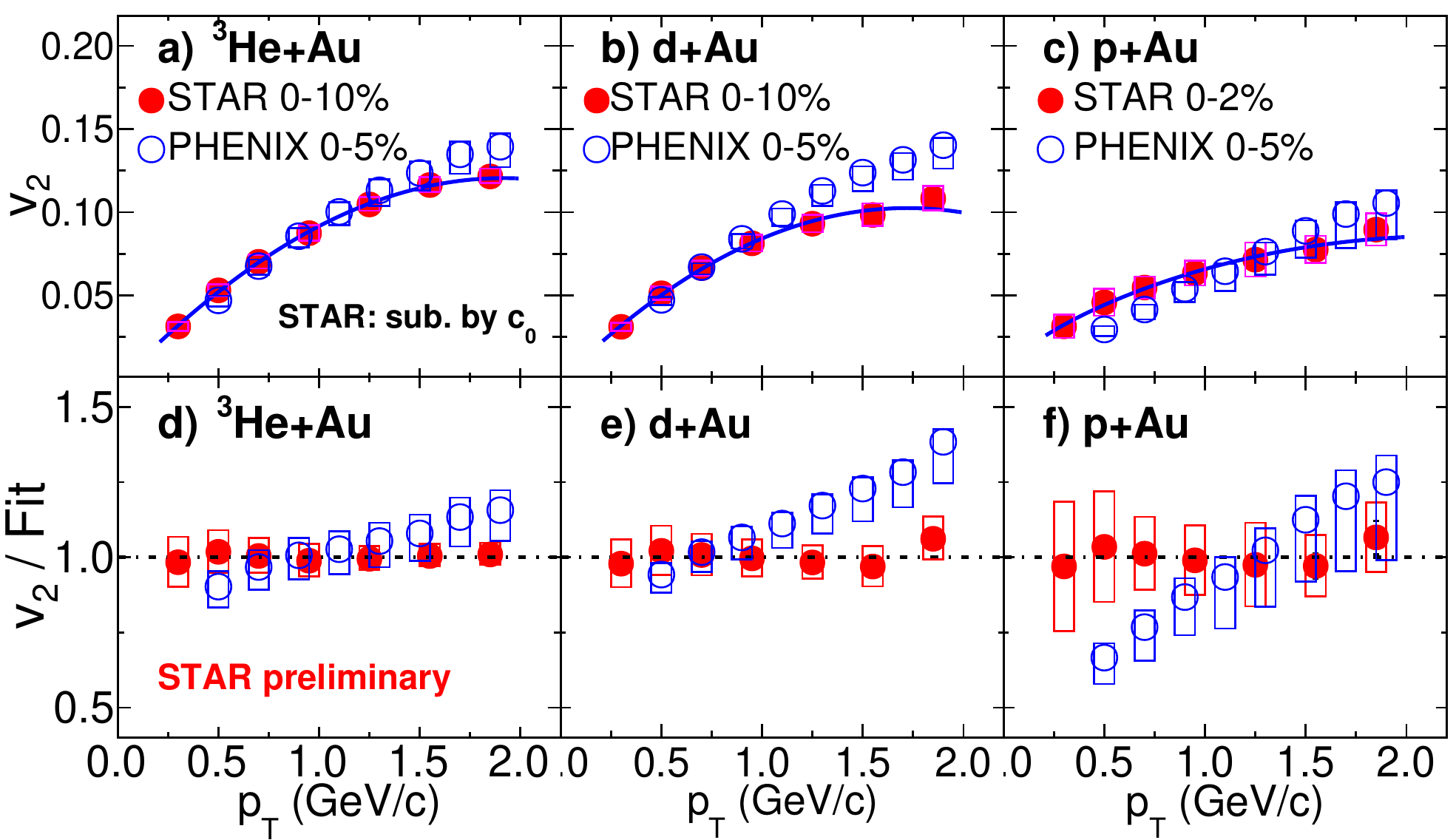}
\includegraphics[width=0.49\linewidth]{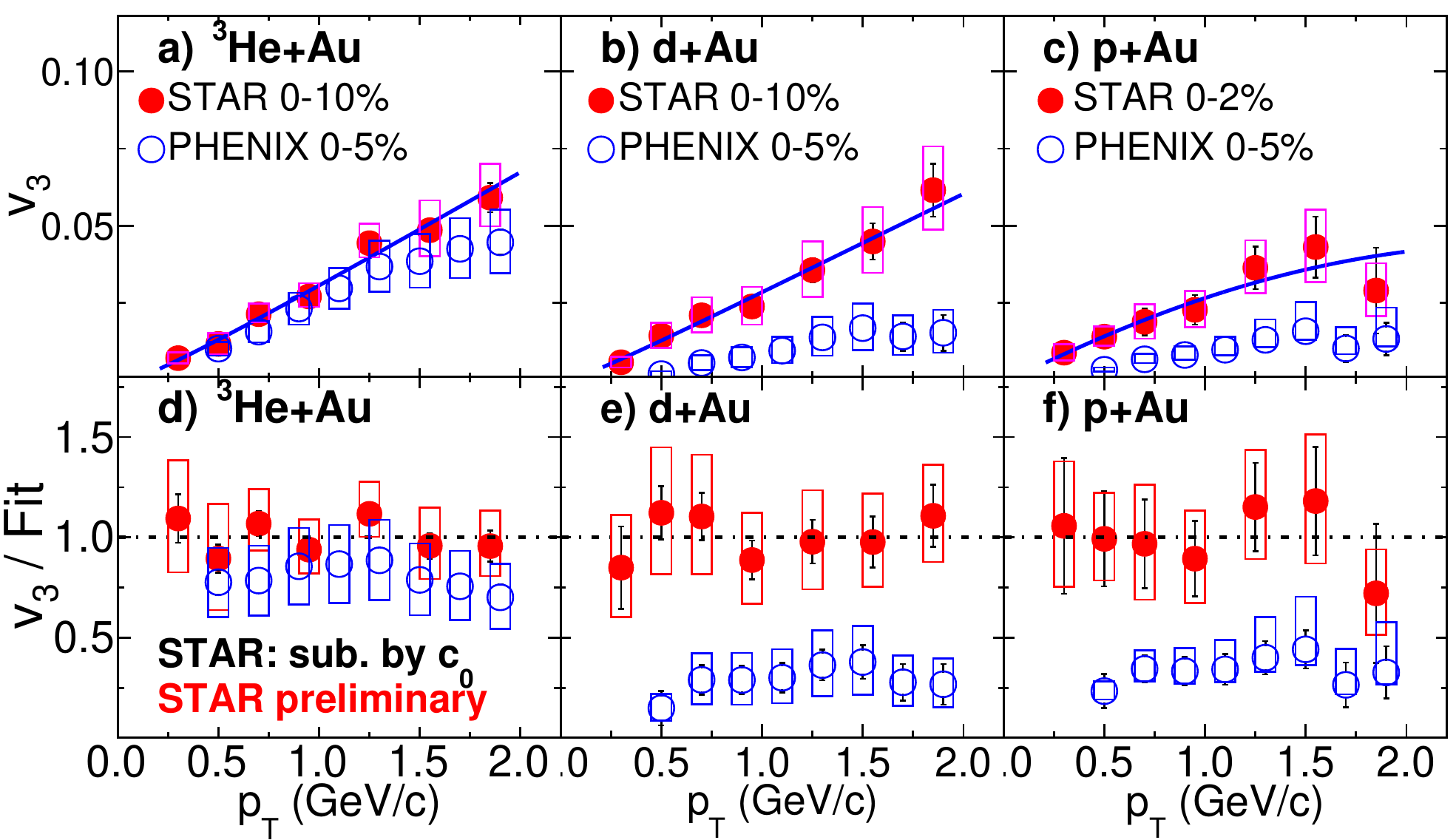}
	\vskip -0.35cm
 \caption{
Comparison of the $v_2(p_T)$ (left panel) and $v_3(p_T)$ (right panel) measurements  obtained by STAR and PHENIX.
The solid lines in the top panels represent a fit to the STAR data. The bottom panels show 
the ratio of the respective data to this fit.
}
  \label{starphenixv2}
%\end{wrapfigure}
\end{figure}
The $v_2(p_T)$ and $v_3(p_T)$ measurements for $p$/$d$/$^3$He+Au are compared to published
PHENIX measurements \cite{PHENIX:2018lia} in Fig.~\ref{starphenixv2}.
The comparisons for $v_2(p_T)$ (left panel) show that, within the indicated uncertainties, 
the $v_2(p_T)$ data from both experiments are in reasonable agreement, albeit with modest 
$p_T$-dependent differences for $p_T > 1$~GeV/$c$ ($d$+Au) and $p_T < 1$~GeV/$c$ ($p$+Au). 
The $v_3(p_T)$ data for $^3$He+Au  (right panel) are also in reasonably good agreement. 
However, the $v_3(p_T)$ measurements for $p$+Au and $d$+Au are about a factor of 3-4 larger than 
those reported by PHENIX, and lie well outside the statistical and systematic uncertainties of the 
current measurements.
The STAR results indicate that the fluctuations-driven $v_3(p_T)$ is system-independent which contrasts with the 
earlier report of a system-dependent $v_3(p_T)$ \cite{PHENIX:2018lia}.
\begin{wrapfigure}{R}{0.34\linewidth}
  %%\begin{center}
	\centering
\includegraphics[width=0.71\linewidth]{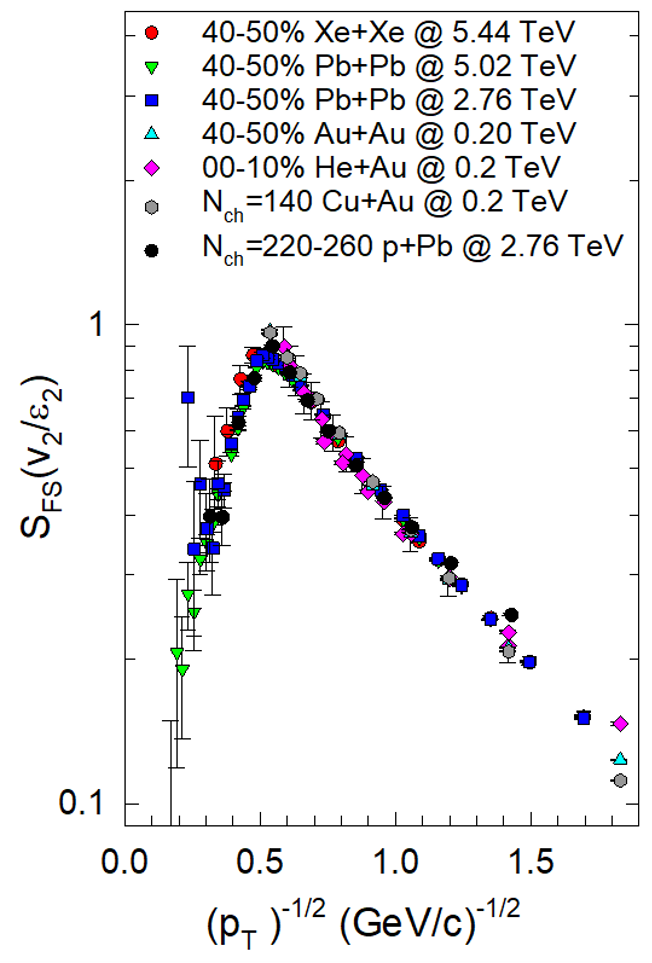}
\vskip -0.35cm
  %%\end{center}
\caption{Anisotropy scaling function for several collision-systems and beam energies.
}
\label{scalingfunction}
\end{wrapfigure}
%\end{figure}
%

The non-flow mitigated $v_n$ measurements shown in Fig.~\ref{fig:method}, can be checked for 
the respective influence of collision-system size ($\rm N_{ch}$), $\varepsilon_n$ and its 
attendant subnucleonic fluctuations and viscous attenuation, via an anisotropy scaling function \cite{Lacey-s}. The scaling 
function $\rm {S_{FS}(v_n/\varepsilon_n, p_T, N_{ch}, \eta/s,\hat{q})}$:
$v_n/\varepsilon_n =  {\exp{( - n[ {n {\beta^{'}} + 
\kappa p_T^2} ]\frac{1}{{(RT)\sqrt{p_T}}}}})$, 
which incorporates the physics of 
jet suppression:
${R_{{\rm{AA}}}}({p_T},L) \simeq \exp [ {\frac{{2{\alpha _s}{C_F}}}{{\sqrt \pi  }}L\sqrt {\mathord{\buildrel{\lower3pt\hbox{$\scriptscriptstyle\frown$}} 
\over q} \frac{\Im }{{{p_T}}}} } ],\frac{{{R_{{\rm{AA}}}}({{90}^0},{p_T})}}{{{R_{{\rm{AA}}}}({0^0},{p_T})}} 
= \frac{{1 - 2{v_2}({p_T})}}{{1 + 2{v_2}({p_T})}}$
and hydrodynamic viscous attenuation: 
${v_n}/{\varepsilon _n} \propto {\exp{( - n[ {n {\beta} + 
\kappa p_T^2} ]\frac{1}{{RT}}}}),RT \propto {\langle {{{\rm{N}}_{{\rm{chg}}}}}\rangle ^{1/3}}$, 
confirms these dependencies via a collapse of diverse measurements of $v_n$ on to a single curve,  
for fully constrained scaling coefficients. 
In turn, the coefficients give insight on the magnitude of the associated transport coefficients.
The scaling function shown in Fig. \ref{scalingfunction}, indicates that the measurements 
are consistent with the final-state (FS) effects which account for 
the broad spectrum of collision-system sizes and energies summarized in the figure. 
Note the jet quenching(viscous attenuation) branch for $p_T > 4$($p_T < 4$)~GeV/$c$.
%Note the jet-quenching and viscous hydrodynamical branches for $p_T\agt 4$~GeV/$c$ and $p_T\alt 4$~GeV/$c$
%respectively.
The resulting scaling coefficients not only suggest an increase in the magnitude of the specific viscosity 
$\left< \eta/s(T, \mu_B) \right>$, from RHIC to LHC energies, but also an increase for relatively small collision-systems.

\section{Summary}
New STAR differential and integral $v_n$ measurements that explicitly account for non-flow contributions, 
are reported  for $p$/$d$/$^3$He+Au collisions at $\sqrt{s_{_{NN}}}=200$~GeV.  
The measurements which are compared to published PHENIX results, indicate system-independent 
values of  $v_2$ and $v_3$ for comparable charged hadron multiplicity, that are consistent with the 
critical influence of both “size”  ($\rm N_{ch}$) and the subnucleonic-fluctuations-driven eccentricities, $\varepsilon_{2,3}$. 
However, they are inconsistent with the notion of shape engineering in $p$/$d$/$^3$He+Au collisions.
The scaling function derived from the measurements, confirm the important role of final-state effects 
across a broad spectrum of collision-system sizes and energies, and suggests 
 an increase in $\eta/s(T,\mu_B)$ for small collision-systems. Future supplemental measurements at RHIC and the LHC,
for systems such as O+O, could provide additional constraints and insights.

%% References


\bibliographystyle{elsarticle-num}
\bibliography{Proc_refs}






% \end{thebibliography}

\end{document}